# FORMATION AND EVOLUTION OF PLANETARY SYSTEMS (FEPS):
## PRIMORDIAL WARM DUST EVOLUTION FROM 3-30 MYR AROUND SUN-LIKE STARS

M.D. Silverstone[1], M.R. Meyer[1], E.E. Mamajek[2], D. C. Hines[3], L.A. Hillenbrand[4], J. Najita[5], I. Pascucci[1], J. Bouwman[6], J.S. Kim[1], J.M. Carpenter[4], J.R. Stauffer[7], D.E. Backman[8], A. Moro-Martin[9], T. Henning[6], S. Wolf[6], T.Y. Brooke[7], D.L. Padgett[7]

ABSTRACT

We present data obtained with the Infrared Array Camera (IRAC) aboard the *Spitzer Space Telescope (Spitzer)* for a sample of 74 young (t < 30 Myr old) Sun-like ($0.7 < M_*/M_\odot < 1.5$) stars. These are a sub-set of the observations that comprise the *Spitzer* Legacy science program entitled the *Formation and Evolution of Planetary Systems* (FEPS). Using IRAC we study the fraction of young stars that exhibit 3.6-8.0$\mu m$ infrared emission in excess of that expected from the stellar photosphere, as a function of age from 3-30 Myr. The most straightforward interpretation of such excess emission is the presence of hot (300-1000K) dust in the inner regions (< 3 AU) of a circumstellar disk. Five out of the 74 young stars show a strong infrared excess, four of which have estimated ages of 3-10 Myr. While we detect excesses from 5 optically thick disks, and photospheric emission from the remainder of our sample, we do not detect any excess emission from *optically thin disks* at these wavelengths. We compare our results with accretion disk fractions detected in previous studies, and use the ensemble results to place additional constraints on the dissipation timescales for optically-thick, primordial disks.

*Subject headings:* circumstellar matter — infrared: stars — planetary systems: protoplanetary disks

[1] Steward Observatory, The University of Arizona, Tucson, AZ
[2] Harvard Smithsonian Center for Astrophysics, Cambridge, MA
[3] Space Science Institute, Boulder, CO
[4] Astronomy, California Institute of Technology, Pasadena, CA
[5] National Optical Astronomy Observatory, Tucson, AZ
[6] Max-Plank-Institut für Astronomie, Heidelberg, Germany
[7] Spitzer Science Center, Caltech, Pasadena, CA
[8] NASA-Ames Research Center, Moffett Field, CA
[9] Astrophysical Sciences, Princeton University, Princeton, NJ

## 1. INTRODUCTION

Low mass stars form surrounded by circumstellar accretion disks that contain the raw material for planet formation (e.g. Hillenbrand et al. 1998). These primordial accretion disks are initially comprised of gas and dust in the interstellar mass ratio (~100) with the opacity dominated by dust grains with temperatures ranging from T≈1400K (the dust sublimation temperature) at the smallest radii (e.g., Muzerolle et al. 2003) to 30K at the largest radii (e.g., Beckwith et al. 1990). Recent observations in the optical and near-infrared confirm the trend found by Strom et al. (1989) suggesting that inner dust disks (< 0.1 AU) dissipate, or grains agglomerate into larger particles, on timescales of 1–10 Myr, with 50% of low mass stars losing their inner dust disks within 3 Myr (Haisch et al. 2001; Hillenbrand et al., in preparation). The loss of inner dust disks appears to be correlated with the cessation of material accreting from the disk onto the central star (Gullbring et al. 1998). Observations at longer wavelengths ($\lambda \geq 10\mu m$) that trace cooler dust suggest that dust disks at larger radii (0.3–30 AU) dissipate or evolve on timescales comparable to the dust inside 0.1 AU (e.g., Meyer & Beckwith 2000, Mamajek et al. 2004; Carpenter et al. 2005, Andrews & Williams 2005).

The expected timescales for planet formation suggest that it may play a role in the dissipation of primordial accretion disks. Current models of gas-giant planet formation indicate that Jupiter mass planets orbiting at 3–10 AU take millions of years to form, and isotopic evidence from terrestrial and lunar samples indicate that the Earth–Moon system was 80–90% complete at an age of 30 Myr (Kleine et al. 2002, 2003). Therefore, the observational constraints on the evolution of disks may provide insight into the formation of planetary systems like our own.

One of the goals of the Formation and Evolution of Planetary Systems (FEPS) *Spitzer Space Telescope (Spitzer)* Legacy Science Program (Meyer et al. 2005a) is to understand the transition from primordial, gas-rich disks containing interstellar dust grains to gas-poor debris disks that contain dust grains generated from the collision of larger bodies. *Spitzer* provides new capabilities to attack this problem, enabling us to probe the disks over wavelength ranges that are difficult or impossible to observe from the ground. In this contribution we present observations of young ($t_* = 3–30$ Myr) Sun-like ($M_* = 0.7–1.5 M_\odot$) stars obtained with the Infrared Array Camera (IRAC), which provides photometry from 3.6–8.0$\mu m$. We use these data to measure the frequency of stars that exhibit 3.6–8.0$\mu m$ emission in excess of that expected from the star alone, and thus to constrain the dissipation of material at moderate temperatures (and orbital distances) on timescales thought to correspond to the epoch of planet-building. In Section 2, we describe the sample, the IRAC observations, and the data reduction. In Section 3, we present the results. In Section 4, we discuss the implications of these results in the context of recent work on the evolution of inner disks around low mass stars and theories of planet formation. Finally in Section 5 we summarize our main conclusions.

## 2. OBSERVATIONS & DATA REDUCTION

FEPS targets were selected based on their ages (spanning 3-3000 Myr), masses (near solar), distances (< 170 pc), infrared background (low), and without bias with respect to previous knowledge of infrared excess properties. FEPS stellar ages were determined by a variety of methods including H-R diagram location and isochrone fitting, Li I ($\lambda$ 6707Å) equivalent width, X-ray luminosities, fractional Ca II H&K emission, rotation, and membership in known clusters and associations (Hillenbrand et al. 2005). There are ~80 stars in the FEPS sample with ages ≤ 30 Myr. Of these, 74 were observed with IRAC before October 5, 2004 and are discussed herein. Of the 74 targets, 48 are members of stellar associations: Scorpio-Centaurus OB Complex subgroups Upper Scorpius (US: t ≈ 5 Myr; Preibisch & Zinnecker 1999), Upper Centaurus-Lupus (UCL: t ≈ 14 Myr; Mamajek et al. 2002), and Lower Centaurus-Crux (LCC: t ≈ 16 Myr; Mamajek et al. 2002). Eight of our targets are associated with two star-formation regions: Corona Australis (CrA; t ≈ 7 Myr) and Chamaeleon I (ChamI; t ≈ 1 Myr); a range of ages have been reported for stars in the vicinity of these cloud complexes (e.g. Neuhaeuser et al. 2000, Alcala et al. 1997), Hillenbrand et al. (2005) assesses the ages of these objects individually based on various age diagnostics. All are consistent with having ages of 3-10 Myr, with little evidence for any of the stars being older 10 Myr. One object has some indication that it might be younger (RX J1111.7-7620), however its inclusion or exclusion has negligible impact on this results from this study. For the remaining 18 stars, ages are determined using a variety of age indicators used above, calibrated with the cluster samples of well-determined age. Age estimates and uncertainties are further detailed in Hillenbrand et al. (2005). To search for evolutionary trends in our young star sample, these 74 targets were divided into two *a priori* determined logarithmic age-bins with widths larger than the estimated uncertainties (~ 50%), for statistical analysis: 29 fall in the 3–10 Myr bin, and 45 fall in the 10–30 Myr bin. Target names, spectral types, and our assigned age bins are reported in Table 2.

The new observations reported here used the 32x32 pixel subarray mode of the IRAC instrument on-board *Spitzer* (Fazio et al. 2004), with a field of view of ~39"x39". Data were obtained in IRAC Channels 1 (3.6$\mu m$), 2 (4.5$\mu m$) and 4 (8.0$\mu m$). The integration times were chosen to yield 2% photometric accuracy on the photospheric flux for each star in all three channels (absolute flux calibration remains the dominant systematic uncertainty). The IRAC effective integration time per frame is 0.01s, 0.08s, or 0.32s for 0.02s, 0.1s, or 0.4s "frame-times," respectively (*Spitzer* Observing Manual Version 4.0). For a given target the frame time is the same for all channels. Within each IRAC channel, the target was placed at each of 4 dither positions separated by ~3-14", corresponding to the "random" dither pattern at the "medium" dither scale. At each of the 4 dither positions, 64 images were taken at the same frame-time. The 4 dithers resulted in 4×64 = 256 images of the object in each channel resulting in total integration times (given in Table 2) of 2.56, 20.48, or 81.92 seconds. Raw data frames were processed through the *Spitzer*

Science Center (SSC) Pipeline version S.10.5.0. to produce Basic Calibrated Data (BCD) image cubes.

Photometry was performed on each of the 256 images using the IDL-based image analysis package IDP3, which was developed by the Instrument Definition Team for the NICMOS instrument on the *Hubble Space Telescope* (Schneider & Stobie 2002) and modified for use with *Spitzer* data. We used a 3-pixel radius target aperture, centered on the location of a Gaussian fit to the stellar image (typical FWHMs ≈ 1.2-2 pixels). The background annulus was 10-32 pixels. This annulus circumscribes the 32x32 array and thus uses all pixels outside of a 10-pixel radius from the target. The background flux in the target aperture was approximated by the median pixel value derived for the background pixels, scaled to the area of the target aperture, and subtracted. An aperture correction was then performed using the values published in the Infrared Array Camera Data Handbook v. 1.0 Table 5.7, for source radius 3.0 pixels and background annulus of 10-20 pixels. No color corrections were applied. The flux densities are given in Table 2.

The distribution of measured fluxes for the 256 images is not Gaussian. We use two methods to estimate the uncertainty. First, we assign the standard deviation of the 256 measurements as the one-sigma, internal uncertainty for each observation. The resulting median values for all 74 targets are 2%, 3%, and 6% for channels 1, 2, and 4, respectively. We consider these values over-estimates of the error in the mean of the measurements; we cannot derive the error in the mean in the standard manner, by dividing the standard deviation of the measurements by the square-root of the number of measurements, due to the non-Gaussian nature of the measurement distribution. Second, we adopt the standard-deviation of the ratio of the predicted photospheric flux and the mean of the observed fluxes for the targets with no apparent excess (see Section 3.1 for the method of photospheric estimation and a discussion of the systems with apparent excess). Using this second method we calculate median uncertainties of 4.5%, 5.2%, and 4.5% for IRAC bands 1, 2, and 4, respectively. We conservatively adopt this second method as the "maximum" typical internal error in the set of observations. As this distribution of ratios contains a component of the error in the estimation of photospheric flux, this should be an over-estimate of the internal errors for the typical observation. The total uncertainties for the measured flux densities are dominated by the 10% calibration uncertainty for all IRAC bands as determined by the SSC[10]. Further information about photometry and noise analyses of FEPS data appear in the explanatory supplement to the FEPS Legacy Science Program Delivery of Enhanced Products (Hines et al. 2005a)[11].

## 3. RESULTS AND ANALYSIS

### 3.1 *Stars with Infrared Excess Emission*

Our goal is to distinguish stars with IRAC flux densities in excess of the expected photospheric emission. To accomplish this we constructed a color-color diagram using the three IRAC measurements for each star plus the Ks-band magnitude from the Two-Micron All Sky Survey (2MASS)[12]. This diagram allows us to determine empirically the locus of points that define "bare" photospheres, and then look for sources with unusually red colors indicative of warm circumstellar dust. Figure 1 shows the Ks-[3.6$\mu$m], [4.5$\mu$m]-[8.0$\mu$m] color-color diagram for the 74 stars in our sample. The typical error, estimated as described above, is plotted as a cross in the upper-left of this figure. Five stars stand out in this diagram as having infrared excess in the IRAC bands all detected at signal-to-noise ratios of 5–19, well-separated to the right and above the cluster of 69 points in the lower-left of Figure 1. *None of the remaining targets has excesses detected at SNR ≥ 3*. In general the IRAC colors of the five stars distinguished with excesses are consistent with those expected from actively accreting classical T Tauri stars (CTTS; Class II objects, e.g. Allen et al. 2004, Bouwman et al. in preparation).

We present the spectral energy distributions (SEDs) of the five stars (RX J1111.7-7620, PDS 66, [PZ99] J161411.0-230536, RX J1842.9-3532, and RX J1852.3-3700) in Figure 2. In addition to the IRAC flux densities we show spectra obtained with the Infrared Spectrograph (IRS: Houck et al. 2004, Bouwman et al. 2005) and photometry from the Multi-Band Imaging Photometer for *Spitzer* (MIPS: Rieke et al. 2004, Kim et al. 2005). We also include IRAS (Moshir et al. 1990; Beichman et al. 1988), ISO (Spangler et al. 2001), and (sub)-millimeter (Carpenter et al. 2005) photometry when available. The observed SEDs are discussed in detail in a forthcoming contribution (Bouwman et al. in preparation) along with models investigating the grain mineralogy. The expected photospheric emission for each star was estimated by fitting Kurucz model atmospheres with convective overshoot to published optical (including, if available, *BV* Johnson, *vby* Stromgren, *BV* Tycho, *H p* Hipparcos, *RI* Cousins) and near-infrared (2MASS *J* and *H*) photometry, with visual extinction treated as a free parameter for targets more distant than 40pc. Additional details concerning the estimation of stellar SEDs for FEPS program stars can be found in Meyer et al. (2004) and Carpenter et al. (2005, in preparation). Brief descriptions of each of these targets and their observed SEDs are now presented.

### 3.2 *Notes on Individual Systems*

#### *RX J1842.9-3532*

RX J1842.9-3532 is a CTTS discovered by Neuhaeuser et al. (2000) in their X-ray survey of the vicinity of the CrA molecular cloud. Neuhaeuser et al. (2000) found strong H$\alpha$ emission (EW = -30.7 Å) and found the star to be very Li-rich like other T Tauri stars (EW = 0.38 Å). They estimate a K2 spectral type, a visual extinction of 1.5 mag, an age of 10 Myr, and a mass of 1.1M$_\odot$ from D'Antona & Mazzitelli (1994) isochrones. RX J1842.9-3532 is also detected in three IRAS bands, with a flux of about 1 Jy at 60$\mu$m. Carpenter et al. (2005) reports a detection at 1.2 mm of 88±11 mJy. From the spectra of Neuhaeuser et al. (2002) Fig. 3 the H$\alpha$ profile is double peaked, providing evidence for active accretion, though the authors do not discuss the H$\alpha$ profile. Although we have classified the two CTTS near the CrA clouds (RX J1852.3-3700 {see below} and RX J1842.9-3532) as kinematic CrA members, we believe that these stars are somewhat older than the TTSs observed near the R CrA molecular cloud core (c.f. Mamajek, & Feigelson, 2001). We estimate an age of 8 Myr (Hillenbrand et al. 2005).

---

[10] http://ssc.spitzer.caltech.edu/irac/calib/overview.html
[11] http://data.spitzer.caltech.edu/popular/feps/20050608_enhanced_v1/Documents/FEPS_Data_Explan_Supp_v2.pdf

[12] The measured flux densities in the three IRAC bands in Table 1 were converted to magnitudes, using the zero-points published in the IRAC Data Handbook v. 1.0 Table 5.1.

The excess of RX J1842.9-3532 begins at wavelengths shorter than 2$\mu$m, indicating the presence of significant quantities of hot dust, and rises at wavelengths longer than 15$\mu$m, suggesting the presence of a disk atmosphere heated by irradiation from the central star as described in the models of Chiang and Goldreich (1997) as well as D'Alessio et al. (1999). The spectrum features a prominent and asymmetric 10$\mu$m feature, indicative of small amorphous silicate grains, as described by Chiang and Goldreich (1997) and D'Alessio et al. (1999) models (optically-thin hot dust above a cooler optically-thick disk mid-plane), with some evidence for crystalline grains.

### RX J1852.3-3700

Neuhaeuser et al. (2000) also identified RX J1852.3-3700 as a Li-rich K3 star with broad H$\alpha$ emission (EW=-33.8 Å), with historical evidence of large variation. Neuhaeuser et al. (2000) conclude that RX J1852.3-3700 is a CTTS. The membership of RX J1852.3-3700 to the R CrA dark cloud is strongly supported by its proper motion and radial velocity, which are very similar to the CrA cloud members (Neuhaeuser et al. 2000). We therefore assign a distance of ~130 pc to RX J1852.3-3700, which is the distance to the CrA star forming region (Marraco & Rydgren, 1981; Casey et al. 1998), and estimate an age of 8 Myr (Hillenbrand et al. 2005).

The excess of RX J1852.3-3700 begins near 3.6$\mu$m (as shown in Figure 1) and sharply rises beyond 15$\mu$m. This may suggest the presence of a large inner hole in an optically thick disk similar to that inferred for TW Hya and CoKu Tau/4 (Uchida et al. 2004; Forrest et al. 2004), and a flared outer disk. The spectrum features a prominent, asymmetric 10$\mu$m feature, suggestive of small amorphous silicate grains with perhaps a modest crystalline component.

### PDS 66

PDS 66, also known as MP Mus, is a bright IRAS source that was first classified as a CTTS by Gregorio-Hetem et al. (1992). Their spectroscopy reveals H$\alpha$ emission with EW = -47 Å and a Li I ($\lambda$ 6707Å) absorption with EW = 0.37 Å. The position and proper motion of PDS 66 suggests that it is a member of LCC, with a secular parallax distance of ~85 pc (Mamajek et al. 2002). Mamajek et al. (2002) formally estimate an age between 7 and 17 Myr using three different evolutionary tracks. We have assigned the mean age of the LCC members (17 Myr; Hillenbrand et al. 2005). Mamajek et al. (2002) also report a spectral type of K1IVe, a visual extinction of 0.17 mag and an X-ray luminosity of 1.6×10$^{30}$ erg/s from the ROSAT All-Sky Survey. The H$\alpha$ profile from Gregorio-Hetem et al. (1992) looks slightly asymmetric but it is difficult to judge. Mamajek et al. (2002) indicate that the spectrum features [O I] ($\lambda$ 6300Å) in emission, another feature typical of CTTS.

The excess of PDS 66 is first evident in the K-band, and rises towards longer wavelengths, as expected for an optically thick dust disk. The spectrum also features a prominent, asymmetric 10$\mu$m feature, indicative of small amorphous silicate grains with some evidence for crystalline grains. Schütz et al. (2005) model the 10$\mu$m feature with two sizes of amorphous and two species of crystalline silicate grains.

### [PZ99] J161411.0-230536

[PZ99] J161411.0-230536 (= GSC 06973-00819, hereafter J161411) is a pre-main sequence member of the Upper Sco association that was discovered by Preibisch et al. (1998), through a spectroscopic survey of X-ray-selected stars. Basic properties of the star are given in Preibisch et al. (1998), Preibisch & Zinnecker (1999), and Mamajek et al. (2004). Its youth is established by strong lithium absorption and a Li I ($\lambda$ 6707Å) equivalent width [EW=0.49Å] within the range typical of T Tauri stars (e.g. Basri et al. 1991). The H$\alpha$ feature has been observed to be nearly filled in with a variable equivalent width of +0.96 (Preibisch et al. 1998) and +0.36, but with blue-shifted emission and red-shifted absorption, indicative of active accretion (Mamajek et al. 2004). We assign the average age of Upper Sco of 5 Myr to this star (Hillenbrand et al. 2005).

An accurate proper motion of J161411 has been measured by the UCAC project ($\mu_\alpha$, $\mu_\delta$ = -12.1±1.6, -23.8±1.9 mas/yr; Zacharias et al. 2004). Using the UCAC2 proper motion, the formalism of deBruijne (1999), and the mean space motion and velocity dispersion for Upper Sco from Madsen et al. (2002), we estimate a membership probability of 95% and a secular parallax distance of 140±14 pc. From this we conclude that J161411 is consistent with being a kinematic member of Upper Sco lying near the mean association distance, 145±2 pc: de Zeeuw et al. (1999).

Metchev et al. (2005) find three candidate companions to J161411 within 5", the brightest and closest is separated from the primary by 0.2" and is 0.5 magnitudes fainter at K-band. The combination of the flux from these two sources likely lead to the anomalous young isochronal age reported by Preibisch & Zinnecker (1999). We therefore argue that a better age estimate for J161411 is ~5Myr, that is the mean isochronal age of the Upper Sco subgroup (Preibisch & Zinnecker 1999).

J161411 was detected by IRAS at 12 and 25$\mu$m and sports a 3.6$\mu$m excess as shown in Figure 2. Mamajek et al. (2004) report mid-IR excess emission towards J161411 and an asymmetry in the H$\alpha$ feature indicating that it is actively accreting from a circumstellar disk. The SED in Figure 2 does exhibit prominent excess emission beginning at or before 3.6$\mu$m and an inflection at ~5$\mu$m. This behavior is reminiscent of that observed by Muzerolle et al. (2003) for a sample of CTTSs in the Taurus dark cloud. They found that the near-infrared colors of objects in their sample could be fit by an inner disk rim with temperatures between 1000-1400 K at radii between 0.07-0.54 AU. The spectrum features a weak 10$\mu$m feature, suggestive of larger amorphous silicate grains and/or dust settling to the midplane of the optically thick disk.

### *RX J1111.7-7620*

RX J1111.7-7620 was first identified as a K3-type weak-line T Tauri star by Walter (1993) based on its H$\alpha$ emission (EW = -7.6 Å) and strong Li I ($\lambda$ 6707Å) absorption (EW = 0.5 Å). The star's proper motion ($\mu_\alpha$, $\mu_\delta$ = -17, +7 ±5, 5 mas/yr) and radial velocity (+16 km/s; Melo 2003) are similar to those of other ChamI members. The star has moderate extinction: $A_V$ = 1.24 mag (Alcala et al. 1997). Using a distance of 150 pc and evolutionary tracks from D'Antona & Mazzitelli (1994), Alcala et al. (1997) determine an approximate age of 2.43±1.08Myr and a mass of 1.17±0.21$M_\odot$. We find that age estimates for this star using H-R diagram location and isochrone fitting, Li I ($\lambda$ 6707Å) equivalent width, and X-ray luminosities range from 2.5–8 Myr, and indicate that this star might be older than the other ChamI stars (Hillenbrand et al. 2005). Speckle observations by Ghez et al. (1997) did not find any close companion to RX J1111.7-7620. However, recent high-resolution spectroscopy by Melo (2003) suggests that it might be a binary. RX J1111.7-7620 is located approximately 25" from a nearby CTTS (XX Cha) and both sources are confused in the IRAS 25 and 60$\mu$m, as well as the ISO 60 and 90$\mu$m beams. The IRAS-detected emission at those wavelengths could be coming from either (or both) system(s).

The excess of RX J1111.7-7620 begins at wavelengths shorter than 2$\mu$m, and is relatively flat out to ~30$\mu$m. The spectrum features a weak 10$\mu$m feature, suggestive of larger amorphous silicate grains and/or dust settling to the midplane of the optically thick disk, with perhaps a small crystalline component.

### 3.3 *Fraction of Systems with Excess*

Only 5 of the 74 systems younger than 30 Myr studied here have infrared excess detected with IRAC. To study the evolutionary behavior of near-infrared excess in systems from 3 to 30 Myr, we compare the fraction of systems observed with detected excess in two different age bins. Four systems from a sample of 29 with ages between 3 and 10 Myr exhibit IRAC excess. Only one of the 45 targets with ages between 10 and 30 Myr features such excess in the IRAC bands. Though the 5 systems with excess emission detected here also exhibit excesses detected by IRAS at longer wavelengths (see Figure 2), as emphasized above, this sample of 74 targets was chosen from a parent sample of over 600 targets independent of knowledge about previous far-infrared measurements. A main result of this investigation is thus optically-thick inner circumstellar disks with temperatures in the range 300-1000K are rare surrounding Sun-like stars at ages greater than 3 Myr old.

The fraction of systems with excess in each of our age bins is presented in Table 1. Based on the number of sources in our sample and assuming Poisson statistics alone, the 1$\sigma$ confidence intervals that the fraction of systems detected here are representative of a parent distribution can be estimated following the method of Gehrels (1986). Although there is a drop in the excess fraction between the 3-10 Myr bin and the 10-30 Myr bin, assuming Poisson statistics dominate the uncertainty, the statistical significance of the drop is about 1$\sigma$. While the widths of our chosen age bins are larger than the estimated uncertainties, it is possible that some targets with ages near the bin boundary belong in an adjacent bin. Thus the Poisson statistics assumed here are potentially underestimates of the true uncertainties.

### 3.4 *Dust Mass limits in Sources Lacking Excess Emission*

Having found only a small fraction of our <30 Myr old sample with measurable infrared excess in the *Spitzer* IRAC bands, how can we interpret our observations of stars that do NOT exhibit evidence for an infrared excess measurable by IRAC? In other words, what limits can we place on the amount of remnant dust surrounding stars lacking obvious excess emission, given our uncertainties? We assume that sources without optically thick disks at IRAC wavelengths could still have undetected optically thin infrared excess emission that is small in magnitude, but within the observational errors. Following Mamajek et al. (2004), we estimate masses corresponding to flux upper-limits for an optically thin disk, whose emission would be at a flux level corresponding to three times the uncertainty of our observations in any of the IRAC bands, as follows. We assume the disk is comprised of single-sized grains, set to the mean particle radius of an equilibrium size distribution (dN/da $\propto a^{-3.5}$), whose minimum is set by the radiation-pressure blow-out size (radius ~0.4$\mu$m for the luminosities of these targets, assuming no gas content in the disk). The dust grains are spread from distances from the central star corresponding to 1400K, approximately the expected dust sublimation temperature for typical dust compositions at the inner rims of accretion disks (Muzerolle et al. 2003), to twice the radii corresponding to 280K, the temperature at which the peak of the emitted spectrum corresponds to the middle wavelength of the 8.0$\mu$m filter. We adapt the radius-temperature relationship for grain sizes which are intermediate between the peak wavelength of the incoming stellar radiation spectrum (i.e. ~0.5–0.7$\mu$m) and the peak wavelength of the grain thermal emission spectrum (i.e. 2-10$\mu$m), i.e. equation 5 of Backman and Paresce (1993). Thus our hypothetical disk extends from ~0.03–3 AU for a typical target. We fit the dust disk model from Mamajek et al (2004), which assumes $\varepsilon_\lambda$ = 1 for $\lambda < 2\pi a$ and $\varepsilon_\lambda = (\lambda/2\pi a)^{-\beta}$, $\beta$ = 1.5 for $\lambda > 2\pi a$, and a constant mass surface density consistent with but not requiring P-R drag dominated dust removal. We set the mass limit of the disk at these radii to that which would produce a flux equal to 3 times the adopted uncertainties of our IRAC observations. We find that the estimated uncertainty for a typical IRAC observation of these FEPS targets corresponds to an upper limit of ~ 4×10$^{-7}$ $M_\oplus$. The range in the upper-limits to excess flux implies a range of dust mass limits from 2.3×10$^{-7}$–3.3×10$^{-6}$ $M_\oplus$. For reference, the portion of the minimum mass solar nebula (Weidenschilling 1977) corresponding to Mercury, Venus, the Earth, and Mars is about 600 $M_\oplus$ (or about 6 $M_\oplus$ of dust assuming a gas-to-dust ratio of 100). For comparison, typical disk masses inferred from sub-mm continuum observations of young (< 1–3 Myr) stars with optically thick inner disks suggest total gas+dust masses of $3\times10^2$–$3\times10^4 M_\oplus$ (Beckwith et al. 1990; Carpenter 2002; Eisner & Carpenter 2003) distributed from ~ 0.1 to ~30 AU. The mass of our current zodiacal dust cloud is estimated to be ~ $3\times10^{-10} M_\oplus$ (Hahn et al. 2002), thus the median upper limit to excess emission in our sample would correspond to 1300 times this mass. It should be noted that this calculation applies to the terrestrial planet forming zone only. Significant amounts of colder dust, e.g. in orbit beyond ~1 AU, would emit at longer wavelengths and would not be detectable with IRAC. Examples of such systems that have been detected in the FEPS project are discussed in Bouwman et al. (2005), Hines et al. (2005b) and Kim et al. (2005).

## 4. DISCUSSION

### 4.1 Comparison to Previous Work

Previous studies have found that the fraction of sources with infrared excesses declines from a high fraction at young ages (>80% at 1 Myr at L-band: Haisch et al. 2001 and N-band: Kenyon & Hartmann 1995) to much lower fractions at ages of 5-10 Myr (e.g., ~10% at L for the 5 Myr old NGC2362: Haisch et al. 2001; 15-20% at N for the ~8 Myr old TW Hya association: Jayawardhana et al. 1999, Weinberger et al. 2003; and the ~10 Myr old $\beta$-Pic moving group: Weinberger et al. 2004). The L-band excess emission indicates the presence of dust located inside of ~0.1 AU, while N-band excess probes dust in a passive reprocessing disk at radii $\leq 1$ AU. The excess fraction of ~10% (14% +11% / -6.6%) measured here with IRAC from 3.5–8.0$\mu$m for sources in the 3-10 Myr age bin is consistent with the range of values observed based on L- or N-band excess.

In Taurus there is nearly a 1:1 correspondence between the presence of an L–band excess and spectroscopic signatures of gas accretion from the inner disk onto the stellar surface (Hartigan, Edwards, and Ghandour, 1995). Based on the detection statistics and the mean age of the population studied, several groups have suggested that the transition from optically-thick active accretion disk to optically-thin passive disk (inside 0.1 AU) occurs rapidly, << 1 Myr (Skrutskie et al. 1990; Kenyon and Hartmann, 1995; Wolk and Walter 1996). Four of the 5 systems with inner disks detected here also show signs of active accretion. While RX J1111.7-7620 shows no strong spectroscopic signatures of active accretion, it does possess a disk that extends to the dust sublimation radius, and its accretion may be episodic.

Excess fractions have also been measured for a few populations older than 10 Myr. Mamajek et al. (2004) detected no N-band excesses among the sources in the ~30 Myr old Tucana-Horologium association, whereas Young et al. (2004) measured an excess fraction of <7% for the comparably aged (30 Myr old) cluster NGC 2547 based on IRAC data. The excess fraction of 2.2% +5.4%/ -1.8% measured for these sources here in the 10-30 Myr age bin is larger than 0, but consistent with these measurements.

It is also interesting to compare the excess fraction measured for the Sco-Cen sources studied here with previous measurements for this population. In their survey of UCL and LCC (17-23 Myr old), Mamajek, Meyer, & Liebert (2002) found a low excess fraction at K : only one of 110 stars exhibited a K-band excess. In contrast Chen et al. (2005) have found a much larger excess fraction (35%) in their MIPS 24$\mu$m survey of Sco-Cen. This difference in the excess fraction may arise in part because the Chen et al. sample contains some hotter, more massive stars, and is sensitive to dust at cooler temperatures than either our, or the Mamajek, Meyer, & Liebert (2002) survey. Our sample, which includes 48 members of the Sco-Cen association (14 in Upper Sco from 3–10 Myr old and 34 in UCL/LCC at 10–30 Myr old) and spans masses from 0.7–1.5$M_\odot$, exhibits an excess fraction of ~7% of all 74 targets in the 3-30 Myr age range (4% of the Sco-Cen members alone) tracing material between 0.1–1 AU. The difference between our results and the Chen et al. (2005) result may arise because disks dissipate from the inside-out and therefore excesses at longer wavelengths (probing larger radii) persist longer than excesses at shorter wavelengths.

### 4.2 Possible Implications

Our observations also place constraints on the generation of dust through collisions of planetesimals in young inner debris belts. We can make a rough estimate of the amount of dust in small particles in the early history of our solar system by considering the amount of unconsolidated material which remains to accrete onto the terrestrial planets. As the terrestrial planets are ~90% complete by 30 Myr (Kleine et al. 2002, 2003), we posit the mass of the remnant debris swarm in the terrestrial planet zone is comparable to the minimum unconsolidated material (10% of ~ 2 $M_\oplus$ = 0.2 $M_\oplus$) located between 0.03-3 AU, which approximates the range of radii probed by our observations (see Section 3.4). We further assume that this mass would be distributed following a collisionally-maintained size distribution $dN/da \propto a^{-3.5}$ extending from a minimum size of roughly 0.5$\mu$m radius (set by radiation pressure "blowout") to a maximum size in the km range (the size of largest objects in planetesimal swarm from which the terrestrial planets were formed; Kenyon & Bromley, 2004). The portion of the total mass of planetary debris in particles distributed between ~0.4-4$\mu$m radius, approximately the range accessible to our observations, would be a few $\times 10^{-6} M_\oplus$. We note that this estimate is probably a very conservative lower limit, as the solid portion of the minimum mass solar nebula contains approximately three times the mass in the inner solar system that eventually ended up in the terrestrial planets (Weidenschilling, 1977). The fact that the observations reported here can constrain the amount of remnant dust in our target systems without detectable excess to $2.3\times10^{-7}$–$3.3\times10^{-6} M_\oplus$ suggests a dust generation rate *lower* than that in our own solar system at comparable epochs (10–30 Myr) if this simple model is indicative.

Despite the plausibility of this extrapolation, in truth, the early evolution of our own solar system is very uncertain. The opacity of inner circumstellar dust disks is expected to drop over time due to grain growth (Beckwith et al. 2000) and possible planet formation, as well as viscous dissipation of the inner disk (Hollenbach et al. 2000). Given the wide range of initial disk masses (Beckwith et al. 1990) and the equally wide range of stellar accretion rates at any given age (e.g. Calvet et al. 2005), one might expect that the epoch at which a system transitions to an optically thin inner disk would vary widely from source to source. Consequently, we might expect to detect a range of excess strengths ranging from bare photospheres to T-Tauri-type excesses. Instead, we find systems populating these two extremes but detect no intermediate excess sources. As Figure 1 demonstrates, 5 optically thick systems were detected at large signal-to-noise ratios, and none of the remaining systems was detected at moderate ratios, which would be expected from optically thin emission from less-massive dust disks.

Perhaps this bimodal distribution reflects a very rapid transition from optically-thick inner gas accretion disks to optically-thin outer passive disks (cf. Skrutskie et al. 1990), which is simply related to the viscous evolution timescale required to clear the inner disk once the outer gas is removed (Hollenbach et al. 2000). Clarke et al. (2001) model the dissipation of the inner disk via viscous accretion and determine that it occurs on a ~15 Myr timescale if photo-evaporation dominates the removal of the outer disk. The excess sources that we have detected show uniformly large excesses, that are often accompanied by accretion signatures (e.g. H$\alpha$ emission, [OI] emission); compared to the rest of the sources in our sample, the accretion phase of these sources appears unusually long-lived. This suggests that they represent the tail-end of a distribution of accretion lifetimes that may be in the process of dissipating via the Clarke et al. mechanism. However, the small number of excess sources in our younger age bin suggests that most inner disks become optically thin via a process that operates more quickly than the Clarke et al., model predicts.

One possibility is that the inner primordial disk is rendered optically thin solely by in situ grain growth and the formation of rocky bodies. Indeed, the formation of planetesimals and larger bodies is believed to occur on timescales << 10 Myr (Weidenschilling & Cuzzi, 1993). If the process of forming these bodies is successful in removing most of the small dust in the vicinity of their orbits, this might explain the low fraction of excess sources that are detected. However, if collisions between these bodies lead to the formation of a ~1000 km size object, significant debris is expected to be generated. Kenyon & Bromley (2004) have predicted that debris generated as part of the process of forming terrestrial planets would be detectable by *Spitzer*. In these models, planetesimal belts are heated dynamically by the presence of larger bodies in the disk, or "self-stirred" due to the emergence of a few ~1000 km "oligarchs" among the planetesimal population. These dynamically active planetesimal belts will collide with large relative velocities generating new dust that is expected to be observable. The Kenyon & Bromley models predict that the N-band excess rises to ~2 magnitudes above the photosphere early on in the formation of a terrestrial planet and drops to 0.25 magnitudes after a period of ~1 Myr. For our typical FEPS source, a 0.25 magnitude excess above the photosphere would correspond to a robust 10-$\sigma$ detection. What limits do our detection statistics place on the frequency of terrestrial planet formation? If all systems undergo a phase of terrestrial planet formation like that described in the Kenyon & Bromley models, we would be able to easily detect the signature of terrestrial planet formation over an interval of 1 Myr after the formation process begins. Assuming this occurs stochastically (e.g., Rieke et al. 2005) with equal probability over the 3-10 Myr age range of our younger age bin, we would expect to detect (1 Myr / 7 Myr) of the 29 sources in the bin or ~4 stars. If terrestrial planet formation is initiated after 10 Myr, but follows the same process described in Kenyon & Bromley (2004), then in our older 10-30 Myr age bin, we would expect to detect (1 Myr / 20 Myr) of the 45 sources or ~2 sources. In comparison, the number of sources detected in these two bins is 4 and 1, similar to the expected number of detections. However, if the dust in these disks is primordial (as suggested by the presence of accretion signatures in 4 of the 5 detected systems) rather than debris generated through terrestrial planet formation, then our detection statistics imply that either: a) a smaller fraction of Sun-like stars (< 25% in the younger age bin) undergoes terrestrial planet formation as described in Kenyon & Bromley (2004); or b) the decay time for the excess produced by terrestrial planet formation is >4 times more rapid than estimated by Kenyon & Bromley (2004). Yet another possibility is that terrestrial planet formation occurs outside the window we have studied, at times <3 Myr or > 30 Myr. In the former case, terrestrial planet formation would appear to occur during the disk accretion phase. This would make it difficult to recognize the debris generated by terrestrial planet formation against the existing excess due to a primordial disk. In the latter case, we would be witnessing the calm before the storm. If >30 Myr is the typical timescale for terrestrial planet formation, it exceeds the timescale inferred for the formation of the Earth (Kleine et al. 2002, 2003).

Finally, we might ask what constraints do these observations place on theories of giant planet formation or the evolution of planetary systems? If infrared excesses are a faithful tracer of the gaseous component of disks, then the relative lack of infrared excess we find at wavelengths < 10$\mu$m would suggest an upper limit of 3-10 Myr on the lifetime of gas-rich inner (< 1 AU) disks. The rarity of gas-rich inner disks beyond an age of 3 Myr would then place strict limits on the time available to form gas giant planets as well as limit migration scenarios for the evolution of their orbits. However, grain growth can significantly alter the continuum opacity of a given column density of material. As a result, the gas-to-dust ratio may vary by orders of magnitude during the planet formation epoch (e.g. Hollenbach et al. 2005, Gorti & Hollenbach, 2004). Therefore, other complementary measurements of the gaseous component of disks are needed to address this issue; this is one of the goals of the FEPS project (Hollenbach et al. 2005).

## 5. SUMMARY

We present initial results from a 3.6-8.0$\mu$m survey for remnant dust in the inner 1 AU surrounding Sun-like stars with ages from 3-30 Myr. These data are a significant fraction (92 %) of those obtained for the youngest stars observed as part of the *Spitzer* Legacy Science Program the *Formation and Evolution of Planetary Systems (Meyer et al. 2005)*. Of our sample of 29 stars with ages 3–10 Myr, only four exhibit evidence for strong infrared excess emission. Among the 45 stars with ages 10–30 Myr, only one exhibits similar infrared excess. Considered together, these five stars have spectral energy distributions consistent with those observed toward well-studied CTTS found in the Taurus dark cloud (e.g. Hartmann et al. 2005). Four ([PZ99] J161411.0-230536, PDS 66, RX J1842.9-3532, and RX J1852.3-3700) of the five stars appear to be actively accreting onto the central star from these inner disks; one (RX J1111.1-7620) is marginally non-accreting. Simple disk models suggest that those stars lacking obvious infrared excess emission in our IRAC data probably have less than $4\times10^{-7} M_\oplus$ of remnant dust in small grains between 0.1-1 AU. Our findings are consistent with previous work indicating that inner disks surrounding T Tauri stars dissipate on timescales of a few million years. We have, however, increased the sample of stars observed with the sensitivity required to detect terrestrial zone dust, and find that most systems are remarkably dust free. These observations may eventually be useful in constraining the frequency and duration of terrestrial planet formation. Further observations are required to distinguish whether or not molecular gas is retained despite the low dust levels in these systems, and/or whether older stars show signs of enhanced dust production in the inner disk due to ensuing collisions of planetesimals.

## 6. ACKNOWLEDGEMENTS

We thank all of the members of the FEPS team for their contributions to the FEPS project and to this study. We would like to thank our colleagues at the *Spitzer* Science Center and members of the instrument teams for their help in analyzing the *Spitzer* data. We also thank an anonymous referee for detailed comments that improved the manuscript significantly. We have used the SIMBAD database. Support for this work, part of the *Spitzer* Space Telescope Legacy Science Program, was provided by NASA through contracts 1224768, 1224634, and 1224566 issued by the Jet Propulsion Laboratory, California Institute of Technology under NASA contract 1407. EEM is supported by a Clay fellowship from the Smithsonian Astrophysical Observatory.

TABLE 1
FRACTION OF SYSTEMS WITH EXCESS

| Age Bin | Total # | Excesses Detected $^{+1\sigma}_{-1\sigma}$ | Fraction $^{+1\sigma}_{-1\sigma}$ |
|---|---|---|---|
| $6.5 \leq \log \text{age} \leq 7.0$ | 29 | $4^{+3}_{-2}$ | $0.14^{+0.11}_{-0.07}$ |
| $7.0 < \log \text{age} \leq 7.5$ | 45 | $1^{+2.4}_{-0.8}$ | $0.022^{+0.054}_{-0.018}$ |

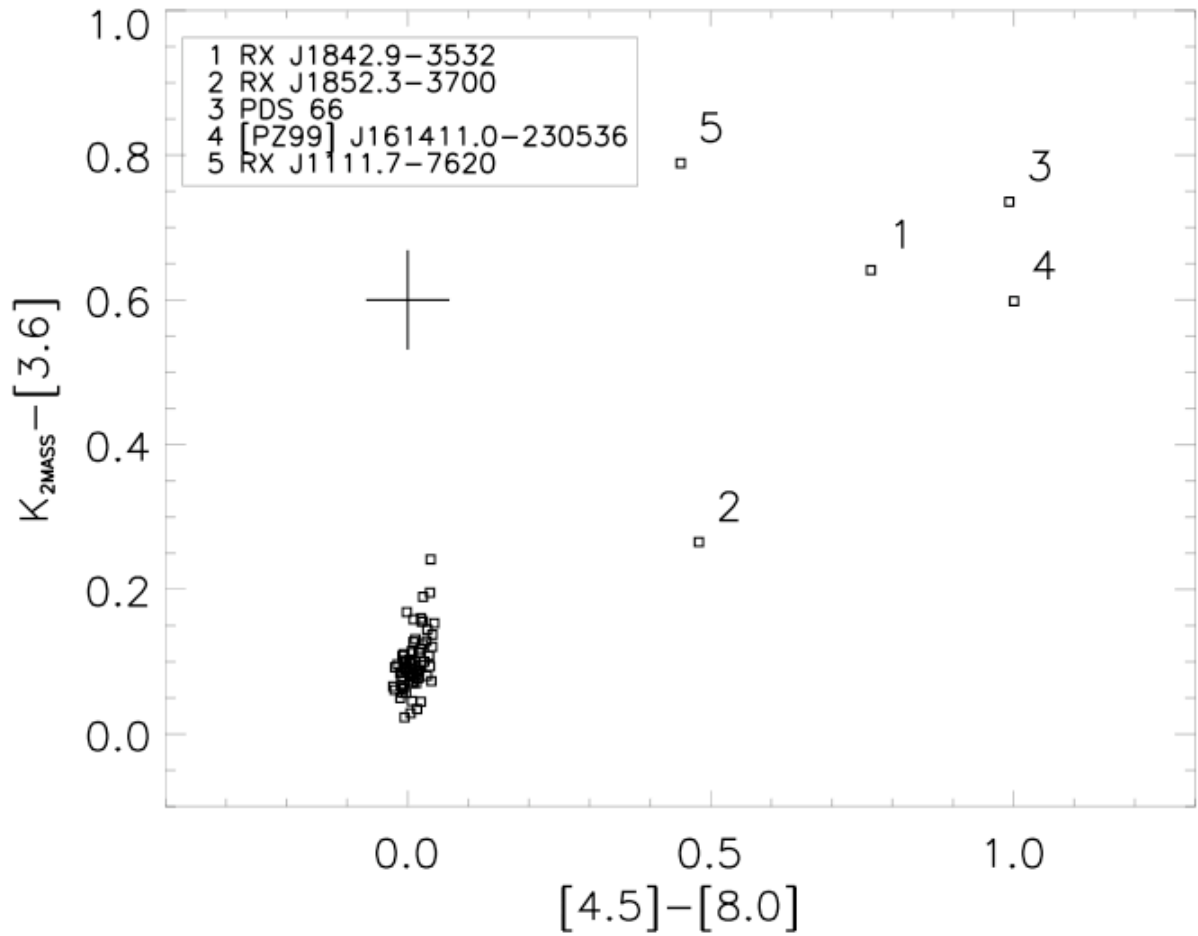

Figure 1. The 2MASS Ks - IRAC 3.6μm, 4.5μm - 8.0μm color-color diagram for the 74 young targets reported in Table 2. Five apparent excess targets appear above and to the right of the cluster of 69 targets with no apparent excess (towards the lower-left portion of the diagram). Plotted below the legend is a typical error, derived as discussed in the text.

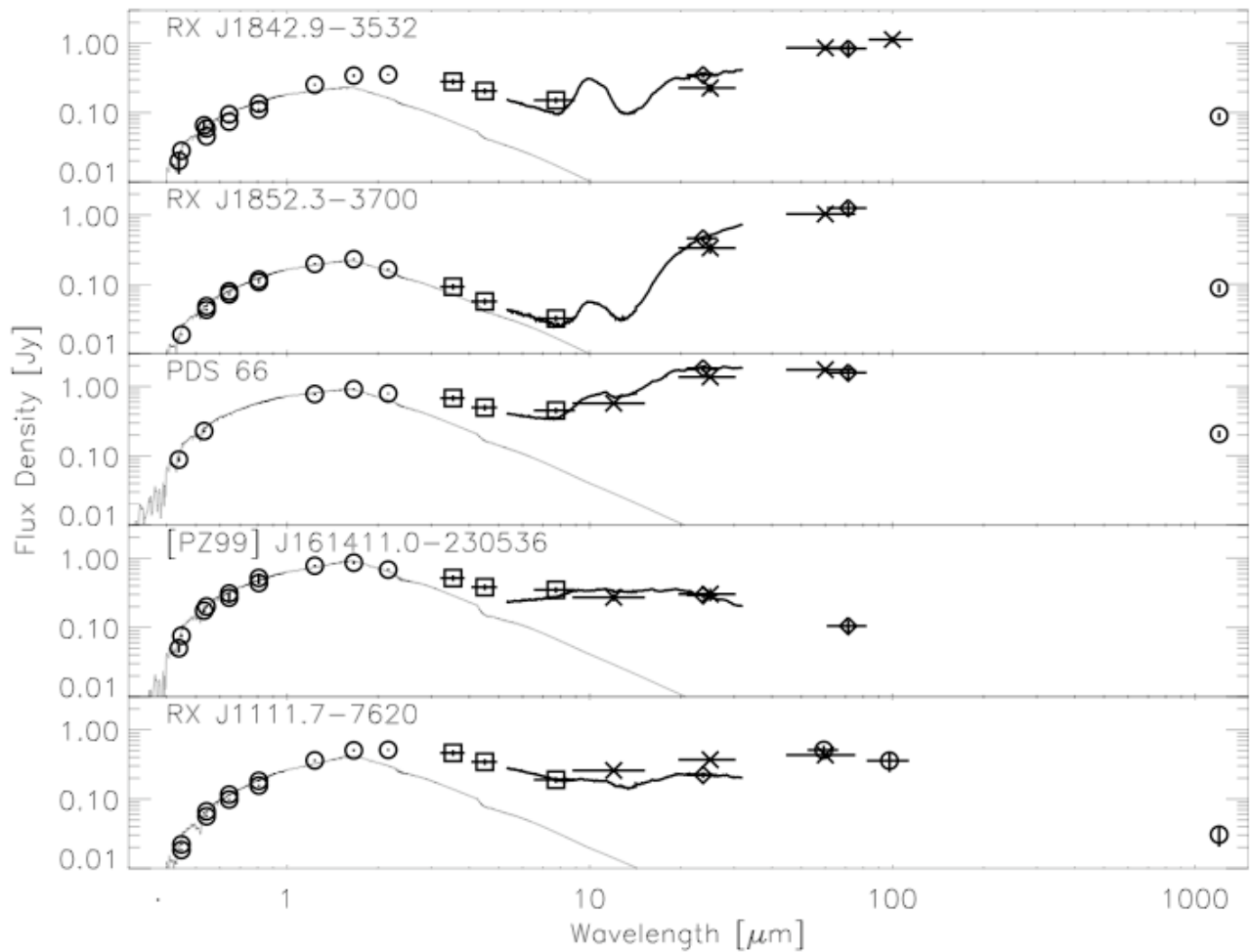

Figure 2. The Spectral Energy Distributions for the five targets with IRAC excesses presented here, featuring Kurucz model photospheres, plotted with solid lines, (blue) fit to optical and near-IR photometry from the literature, plotted with circles (black), including Hipparcos and 2MASS databases. IRAS catalog fluxes are plotted as X's (black). FEPS observations with IRAC at 3.6$\mu$m, 4.5$\mu$m, and 8.0$\mu$m are plotted with boxes (green), IRS low-resolution spectra with a solid line (red) and MIPS 24 and 70$\mu$m photometry are plotted with diamonds (green). ISO 60 and 100$\mu$m fluxes from Spangler et al. (2001) are plotted for RX J1111.7-7600 with circles (blue). Millimeter fluxes from Carpenter et al. (2005) are plotted with circles (green) at the far right of the diagrams. The IRAS, ISO, and MIPS 70$\mu$m measurements of RX J1111.7-7600 are confused with flux from XX Cha, a CTTS ~25" away. In several cases, multiple discrepant observations in the blue likely indicate variation, common to systems with active accretion disks. Such variability is typically smaller in the infrared. The Kurucz model photospheres chosen minimize the $\chi^2$ statistic, taking the measurement uncertainties into account.

TABLE 2
FEPS YOUNG TARGETS AND *SPITZER*/IRAC FLUXES
Targets are listed in RA order

| Source | Spectral Type | Age Bin (Myr) | S(3.6μm) (mJy) | S(4.5μm) (mJy) | S(8.0μm) (mJy) | Integration Time (sec) | Group |
|---|---|---|---|---|---|---|---|
| RE J0137+18A | K3Ve | 3–10 | 678.2 | 395.1 | 141.9 | 2.56 | Field |
| RX J0331.1+0713 | K4(V)/E | 3–10 | 184.1 | 108.5 | 39.0 | 20.48 | Field |
| HD 285281 | K1 | 10–30 | 288.5 | 169.5 | 60.6 | 20.48 | Field |
| HD 285372 | K3(V) | 3–10 | 96.7 | 56.7 | 20.1 | 20.48 | Field |
| HD 284135 | G3(V) | 3–10 | 246.1 | 144.6 | 50.0 | 20.48 | Field |
| HD 284266 | K0(V) | 10–30 | 113.8 | 67.5 | 23.5 | 20.48 | Field |
| HD 279788 | G5V | 3–10 | 122.4 | 71.8 | 25.3 | 20.48 | Field |
| 1RXS J043243.2-152003 | G4V | 3–10 | 110.6 | 64.6 | 22.7 | 20.48 | Field |
| RX J0434.3+0226 | K4e | 10–30 | 50.1 | 29.9 | 10.6 | 81.92 | Field |
| RX J0442.5+0906 | G5(V) | 10–30 | 69.5 | 40.1 | 14.3 | 81.92 | Field |
| HD 286179 | G3(V) | 10–30 | 128.2 | 75.4 | 26.2 | 20.48 | Field |
| HD 286264 | K2IV | 10–30 | 277.3 | 161.5 | 58.2 | 20.48 | Field |
| AO Men | K(3)(V) | 10–30 | 590.2 | 348.9 | 122.1 | 2.56 | Field |
| RX J0850.1-7554 | G5 | 10–30 | 102.5 | 59.6 | 21.1 | 20.48 | Field |
| MML 1 | K1+IV | 10–30 | 237.8 | 137.5 | 49.2 | 20.48 | LCC |
| RX J1111.7-7620 | K1 | 3–10 | 462.6 | 350.2 | 186.0 | 2.56 | ChamI |
| HD 104467 | G5III/IV | 3–10 | 553.7 | 326.0 | 112.1 | 2.56 | Field |
| MML 8 | K0+IV | 10–30 | 172.8 | 101.4 | 36.1 | 20.48 | LCC |
| MML 9 | G9IV | 10–30 | 171.7 | 100.0 | 35.6 | 20.48 | LCC |
| HD 106772 | G2III/IV | 10–30 | 1031.1 | 611.4 | 216.8 | 2.56 | Field |
| HD 107441 | G1.5IV | 10–30 | 264.1 | 157.0 | 54.8 | 20.48 | LCC |
| MML 17 | G0IV | 10–30 | 236.7 | 140.8 | 49.1 | 20.48 | LCC |
| MML 18 | K0+IV | 10–30 | 153.3 | 88.9 | 32.0 | 20.48 | LCC |
| HD 111170 | G8/K0V | 10–30 | 387.5 | 227.7 | 80.5 | 20.48 | LCC |
| MML 26 | G5IV | 10–30 | 143.3 | 83.6 | 29.2 | 20.48 | LCC |
| MML 28 | K2-IV | 10–30 | 96.7 | 56.3 | 20.0 | 81.92 | LCC |
| MML 32 | G1IV | 10–30 | 137.8 | 81.9 | 28.2 | 20.48 | LCC |
| HD 116099 | G0/3 | 10–30 | 116.1 | 68.4 | 24.0 | 20.48 | LCC |
| PDS 66 | K1IVe | 10–30 | 680.4 | 506.5 | 442.4 | 2.56 | LCC |
| HD 117524 | G2.5IV | 10–30 | 231.5 | 135.6 | 47.9 | 20.48 | LCC |
| MML 36 | K0IV | 10–30 | 224.3 | 131.0 | 46.4 | 20.48 | UCL |
| HD 119269 | G3/5V | 10–30 | 278.7 | 163.9 | 57.4 | 20.48 | LCC |
| MML 38 | G8IVe | 10–30 | 112.0 | 66.2 | 23.2 | 20.48 | UCL |
| HD 120812 | F8/G0V | 10–30 | 208.6 | 123.7 | 43.2 | 20.48 | UCL |
| MML 40 | G9IV | 10–30 | 129.6 | 75.5 | 26.9 | 20.48 | UCL |
| MML 43 | G7IV | 10–30 | 121.1 | 70.7 | 24.7 | 20.48 | UCL |
| HD 126670 | G6/8III/IV | 10–30 | 223.7 | 133.1 | 46.6 | 20.48 | UCL |
| HD 128242 | G3V | 10–30 | 232.2 | 138.7 | 48.4 | 20.48 | UCL |
| RX J1450.4-3507 | K1(IV) | 10–30 | 185.0 | 108.7 | 38.7 | 20.48 | UCL |
| MML 51 | K1IVe | 10–30 | 150.5 | 89.8 | 32.0 | 20.48 | UCL |
| RX J1457.3-3613 | G6IV | 10–30 | 154.6 | 91.8 | 32.0 | 20.48 | UCL |
| RX J1458.6-3541 | K3(IV) | 10–30 | 219.4 | 129.7 | 46.4 | 20.48 | UCL |
| RX J1500.8-4331 | K1(IV) | 10–30 | 99.8 | 59.4 | 21.1 | 81.92 | UCL |
| MML 57 | G1.5IV | 10–30 | 123.3 | 71.7 | 25.0 | 20.48 | UCL |
| RX J1507.2-3505 | K0 | 10–30 | 142.5 | 82.8 | 29.4 | 20.48 | UCL |
| HIP 76477 | G9 | 10–30 | 160.1 | 93.6 | 33.7 | 20.48 | UCL |
| V343 Nor | K0V | 10–30 | 1396.8 | 819.1 | 289.2 | 2.56 | Field |
| HD 139498 | G8(V) | 10–30 | 297.6 | 176.2 | 62.3 | 20.48 | UCL |
| RX J1541.1-2656 | G7 | 3–10 | 85.1 | 49.5 | 17.7 | 81.92 | US |
| HD 140374 | G8V | 10–30 | 231.4 | 135.4 | 47.3 | 20.48 | UCL |
| RX J1545.9-4222 | K1 | 10–30 | 216.9 | 127.5 | 46.1 | 20.48 | UCL |
| HD 141521 | G8V | 10–30 | 255.6 | 148.5 | 53.2 | 20.48 | UCL |
| HD 141943 | G0/2V | 10–30 | 876.6 | 523.5 | 180.4 | 2.56 | Field |
| HD 142361 | G3V | 3–10 | 476.7 | 283.1 | 102.5 | 2.56 | US |
| [PZ99] J155847.8-175800 | K3 | 3–10 | 152.7 | 90.4 | 33.0 | 81.92 | US |
| RX J1600.6-2159 | G9 | 3–10 | 139.9 | 82.4 | 29.6 | 20.48 | US |
| HD 143358 | G1/2V | 10–30 | 180.2 | 107.5 | 37.4 | 20.48 | UCL |
| ScoPMS 21 | K1IV | 3–10 | 125.4 | 73.8 | 26.9 | 20.48 | US |
| ScoPMS 27 | K2IV | 3–10 | 192.6 | 113.2 | 41.1 | 20.48 | US |
| [PZ99] J160814.7-190833 | K2 | 3–10 | 140.1 | 82.1 | 29.7 | 20.48 | US |
| ScoPMS 52 | K0IV | 3–10 | 331.8 | 194.4 | 71.1 | 20.48 | US |
| [PZ99] J161318.6-221248 | G9 | 3–10 | 338.2 | 198.4 | 72.0 | 20.48 | US |
| [PZ99] J161329.3-231106 | K1 | 3–10 | 143.8 | 84.6 | 30.8 | 20.48 | US |
| [PZ99] J161402.1-230101 | G4 | 3–10 | 124.2 | 73.0 | 26.5 | 20.48 | US |
| [PZ99] J161411.0-230536 | K0 | 3–10 | 515.7 | 388.6 | 341.8 | 2.56 | US |
| [PZ99] J161459.2-275023 | G5 | 3–10 | 105.3 | 61.2 | 22.3 | 20.48 | US |
| [PZ99] J161618.0-233947 | G7 | 3–10 | 176.6 | 103.3 | 37.6 | 20.48 | US |
| RX J1839.0-3726 | K1 | 3–10 | 115.9 | 68.5 | 24.2 | 20.48 | CrA |
| RX J1841.8-3525 | G7 | 3–10 | 193.8 | 115.5 | 40.1 | 20.48 | CrA |
| RX J1842.9-3532 | K2 | 3–10 | 279.2 | 209.0 | 148.1 | 20.48 | CrA |
| RX J1844.3-3541 | K5 | 3–10 | 146.2 | 85.6 | 30.3 | 20.48 | CrA |
| RX J1852.3-3700 | K3 | 3–10 | 91.9 | 57.5 | 31.4 | 2.56 | CrA |
| HD 174656 | G6IV | 3–10 | 373.3 | 219.5 | 77.7 | 20.48 | CrA |
| RX J1917.4-3756 | K2 | 3–10 | 329.3 | 192.2 | 68.9 | 20.48 | CrA |